\documentclass{aa}
\usepackage{graphicx,amsmath}
\usepackage{txfonts}

\usepackage{natbib} 
\bibpunct{(}{)}{;}{a}{}{,}
\begin{document}
\title{The formation of sunspot penumbra}
\subtitle{I. Magnetic field properties}

\author{R. Rezaei, N. Bello Gonz$\acute{\rm{a}}$lez, \& R. Schlichenmaier}
\titlerunning{Polarimetry of a forming penumbra}
\authorrunning{Rezaei, Bello Gonz$\acute{\rm{a}}$lez \& Schlichenmaier}  
\institute{Kiepenheuer-Institut f\"ur Sonnenphysik, Sch\"oneckstr. 6, D-79104 Freiburg, Germany\\
\email{[rrezaei, nbello, schliche]@kis.uni-freiburg.de}}
\date{Received 15 June 2011; Accepted 7 October 2011}
\keywords{Sun: sunspot -- Sun: magnetic fields -- Sun: photosphere -- Sun: evolution}

\abstract{} 
{We study the magnetic flux emergence and formation of a sunspot penumbra in the active region NOAA\,11024.}
{We simultaneously observed the Stokes parameters of the photospheric iron lines
at 1089.6\,nm with the TIP and 617.3\,nm with the GFPI spectropolarimeters
along with broad-band images using G-band and \ion{Ca}{\sc ii}\,K filters at the German VTT.
The photospheric magnetic field vector was reconstructed from an inversion of the measured
Stokes profiles. Using the AZAM code, we converted the inclination from line-of-sight (LOS) to the local reference frame (LRF).}
{Individual filaments are resolved in maps of magnetic parameters. The formation of
the penumbra is intimately related to the inclined magnetic field. 
No penumbra forms in areas with strong magnetic field strength and small inclination. Within 4.5\,h observing time, 
the LRF magnetic flux of the penumbra increases from 9.7\,$\times\,10^{20}$ to 18.2\,$\times\,10^{20}$\,Mx, while
the magnetic flux of the umbra remains constant at $\sim 3.8 \times\,10^{20}$\,Mx. 
Magnetic flux in the immediate surroundings is incorporated into the spot, and new flux
is supplied via small flux patches\,(SFPs), which on average have a flux of 2--3\,$\times\,10^{18}$\,Mx.
The spot's flux increase rate of 4.2\,$\times\,10^{16}$\,Mx\,s$^{-1}$ corresponds to the merging of one SFP per minute. 
We also find that during the formation of the spot penumbra: 
$a)$ the maximum magnetic field strength of the umbra does not change,
$b)$ the magnetic neutral line keeps the same position relative to the umbra,
$c)$ the new flux arrives on the emergence side of the spot while the penumbra forms on the opposite side, 
$d)$ the average LRF inclination of the light bridges decreases from 50$^{\circ}$ to 37$^{\circ}$, and 
$e)$ as the penumbra develops, the mean magnetic field strength at the spot border decreases from 1.0 to 0.8 kG.}
{The SFPs associated with elongated granules are the building blocks of structure formation in active regions. During the
sunspot formation, their contribution is comparable to the coalescence of pores.
Besides a set of critical parameters for the magnetic field, a quiet environment in the surroundings 
is important for penumbral formation. 
As remnants of trapped granulation between merging pores, the light bridges 
are found to play a crucial role in the formation process. 
They seem to channel the magnetic flux through the spot during its formation. 
Light bridges are also the locations where the first penumbral filaments form.}

\maketitle

\section{Introduction}
The Sun is a unique laboratory for studying cosmological magnetic fields. 
Active regions (AR) are manifestations of large-scale magnetic field in the solar atmosphere. 
The largest magnetic structures in ARs are sunspots. 
To understand the formation of such magnetic features, one has to 
study the fundamental process of flux emergence \citep{lites_09_ar, kosovichev09}. 
The rise of buoyant magnetic flux tubes is the basic picture of the present understanding of 
these processes. Since its introduction by \cite{parker_55}, there have been many attempts to 
simulate the emergence of buoyant flux tubes in the convection zone  \citep{cale_etal95, abbet_etal00, fang_etal10}. 
As a result of flux emergence, pores and 
other smaller magnetic structures also form in ARs. 
An understanding of their structure and evolution
is a necessity for solving the problem of magnetic field generation in the solar interior \citep{parker_79}. 
The formation of the sunspot penumbra and the associated onset of the Evershed flow
\citep{leka_sku98, yang_etal03} is of particular interest in this respect as summarized in the
reviews by \cite{solanki03r}, \cite{weiss_06}, and \cite{rolf09}.
Recently, numerical simulations try to reproduce the flux-emergence phenomena in the solar 
atmosphere \citep{cheung_etal08, sykora_etal09, tortosa_etal09, cheung_etal10, fang_etal10}.

We succeeded in acquiring a spectropolarimetric data set of a forming penumbra (NOAA 11024) with high
temporal cadence and spatial resolution \citep[][hereafter Paper\,1 \& 2]{rolf_etal_10a,rolf_etal_10b}.  
The image sequences in the G-band and in the \ion{Ca}{ii}{\,K} filters 
show that the penumbra forms in segments at the outer boundary of the protospot on the side
away from to the AR center. As the area of the penumbra increases, the area of the umbra
stays approximately constant (Paper 1). Although \cite{zwaan85, zwaan87} compiled observations suggesting 
that sunspots form out of merging pores, we concluded in Paper\,2 that a fraction of the magnetic flux 
required to form a sunspot emerges in the form of granular-scale bipoles between the two polarities of the AR. 
These bipoles are cospatial with elongated granules as seen in intensity. 
Subsequently the bipole extremes separate and the proper polarity merges with the forming spot,  
therefore the magnetic flux of the spot increases and the penumbra forms.

In this Paper, we quantify the magnetic properties of the elongated granules and 
the spot as it develops a penumbra. 
The Evershed flow and velocities will be addressed in the next paper. 
We present a time series of spectropolarimetric maps that were taken simultaneously with
the imaging data presented in Paper\,1. 
The data sets of our two spectropolarimeters the (visible) GREGOR Fabry-P\'erot 2D-interferometer (GFPI)
and the Tenerife Infrared Polarimeter\,(TIP \,II) are described in Section \ref{sec:obs}. 
The Stokes profiles are inverted to retrieve the physical parameters (cf. Sect.~\ref{sec:inv}). 
A comparison between maps of the two independent spectropolarimeters highlights
the reliability of our measurements. In Sect.~\ref{sec:magnetic} we elaborate on the increase in 
magnetic flux as the protospot transforms into a sunspot. Section~\ref{sec:discussion} summarizes and 
discusses our analysis, while Sect.~\ref{sec:conclusion} presents the conclusions.

\section{Observations}\label{sec:obs}
As discussed in Papers\,1 and \,2, the evolution of the AR\,11024 was observed on 
consecutive days in July 2009 at the German Vacuum Tower Telescope \citep[VTT,][]{vtt}. 
From SoHO/MDI images \citep{soho1}, it is seen that two pores emerged on July 3 
and formed a protospot~\citep{rolf_etal_10c}.
In this contribution, we focus on the July 4 observations, during which the leading spot 
of the AR, located at a heliocentric angle $\theta\,\sim$\,28$^{\circ}$, developed a penumbra.

\begin{table}
\begin{center}
\caption{Atomic properties of the observed spectral lines~\citep{nave_etal94}.}
\label{tab:lines}
\begin{tabular}{c c c c c} 
\hline
Line & $\lambda$\,(nm) & Exc pot\,(eV) & $\log(gf)$ & $g$-effective\\\hline
Fe\,{\sc i} & 1089.63 & 3.071 & -2.85 & 1.50\\
Fe\,{\sc i} & 617.334 & 2.223 & -2.88 & 2.50\\
\hline
\end{tabular}
\end{center}
\end{table}

This observing campaign was based on a multi-instrument, multiwavelength optical 
setup like the ones reported by \cite{beck_mikurda_etal_07} for TIP/TESOS \citep[see also ][]{kucera_etal_08}. 
However, this was the first simultaneous TIP/GFPI observing campaign. 
We combined the TIP\,II \citep{collados_etal07} 
attached to the Echelle spectrograph, the  GFPI \citep{puschmann_etal06, nazaret_kneer08}, 
and two speckle-imaging channels, in the G-band and \ion{Ca}{ii}{\,K}.  
A dichroic beam-splitter was used to feed GFPI with visible light and TIP with infrared.
A small fraction of the visible beam was extracted for both imaging systems. 
The scanning mirror of the Kiepenheuer Adaptive Optics System \citep[KAOS,][]{luhe_etal_03} was
used for spatial scanning by shifting the solar image at the entrance of
the spectrograph slit.  Thus, one could built maps of a certain area of the Sun with the slit
spectrograph while scanning in wavelength with the 2D spectrometer and imaging in the speckle channels. 
The filtergrams in the G-band and \ion{Ca}{ii}{\,K} were reconstructed using the KISIP 
code \citep{kisip_2008,woeger_etal08}, achieving a spatial resolution better than 0\farcs3. 
In Paper\,1, properties of the sunspot formation as seen in the reconstructed
image sequences have been presented. Two movies are available as online material in Paper\,1.

\subsection{One-dimensional spectropolarimetric data}
TIP recorded maps in the \ion{Fe}{i} line at 1089.6\,nm (Table\,\ref{tab:lines}). 
The line forms only in a small height range, such that an analysis that assumes that magnetic field 
and velocity are constant across the formation range is more reliable than, e.g., for \ion{Si}{i} 1082.7\,nm which 
has contributions from the entire photosphere. 
Two full sunspot maps were recorded between 10:42--10:58 and 11:43--11:59\,UT 
with a slit width corresponding to 0\farcs3 and a step size of 0\farcs3, a 
scanning range of $24^{\prime\prime}$ (i.e., 80 steps) and a slit length of 78\arcsec. 
During the rest of the observing time, we performed scans with a range of
only $2\farcs1$ to keep the spot within the field of view\,(FOV) of the GFPI, which recorded data 
continuously. The spatial sampling along the slit after binning amounts to $0\farcs35$. 
The spectral sampling is $1.1$\,pm, and the effective exposure time per slit position was 10\,s. 
The polarimetric calibration was performed using the VTT telescope model and the near-infrared
instrumental calibration unit \citep{beck05b}. 
The residual crosstalk was corrected for using the statistical method that is 
described in \cite{schliche_collados02}. 
The 1$\sigma$ noise level in the TIP data is about 1\,$\times$\,10$^{-3}\,I_{\mathrm c}$.

\begin{figure}
\resizebox{8.0cm}{!}{\includegraphics*{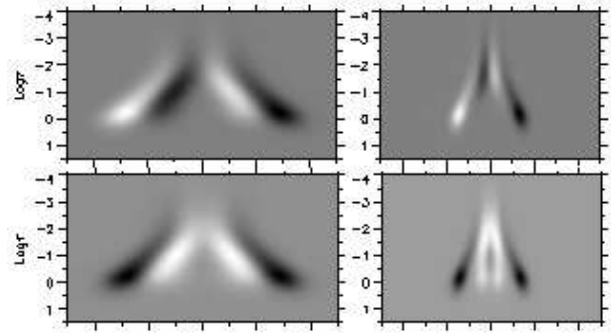}}
\caption{Response of $V(\lambda)$ to LOS velocity (upper row) and magnetic field strength 
(lower row) for Fe\,{\sc i}\,1089.6\,nm (left column) and Fe\,{\sc i}\,617.3\,nm (right column). 
Each tick mark in the $x$-axis corresponds to 10\,pm.}
\label{fig:response}
\end{figure}

\begin{figure*}
\centering
\resizebox{\hsize}{!}{\includegraphics{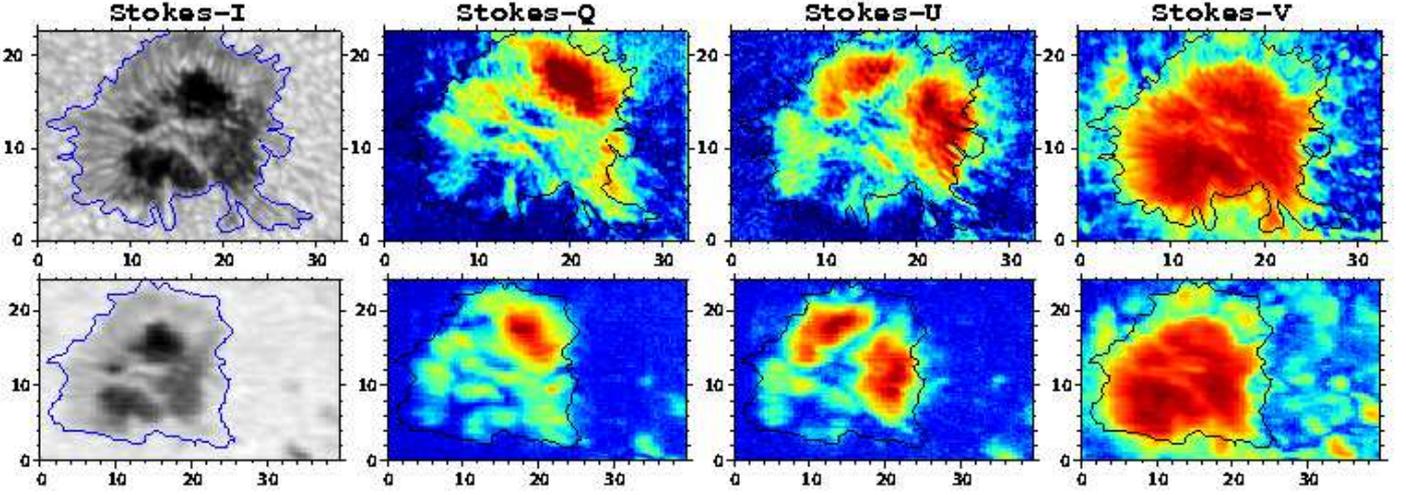}}
\caption{Quasi-simultaneous polarimetric maps of GFPI (Fe\,{\sc i}\,617.3\,nm, \emph{top}) 
and TIP (Fe\,{\sc i}\,1089.6\,nm, \emph{bottom}). \emph{From left to right:} Maps of continuum intensity, 
$I(\lambda_{\mathrm c})$, and the polarimetric 
states $\int\!\mid\!X(\lambda)\!\mid {\rm d}\lambda, X\!\in\!\{Q,\,U,\,V\}$.
Contours show the corresponding continuum boundary of the spot (manually drawn). 
The TIP maps should to be rotated by 3$^{\circ}$ in clockwise direction 
to be aligned with GFPI maps. The GFPI data was acquired at 11:54, while the TIP map 
was scanned from 11:43 until 11:59\,UT (from bottom to top). Tick marks are in arcsec.}
\label{fig:compare}
\end{figure*}
\begin{figure*}
\centering
\resizebox{\hsize}{!}{\includegraphics*{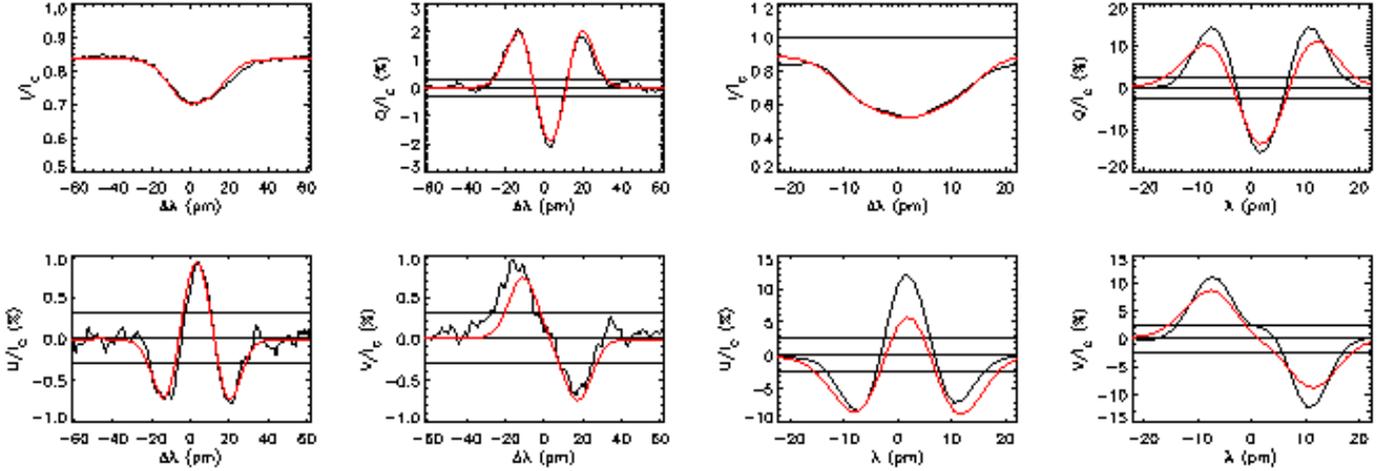}}
\caption{Example of inversion results. The black and red curves show the observed and inverted profiles, respectively. 
The two sets of profiles correspond to the limb-side penumbra.
\emph{Left:} TIP, \emph{right:} GFPI. The horizontal lines in $Q,U,V$ panels mark the zero and $\pm3\,\sigma$ noise levels. 
The GFPI profiles are filtered (cf. Sect.\ref{sec:twodim}).
The selected inversion setup only produces symmetric profiles, 
causing the imperfect fit of the e.g., Stokes-$I$ and $V$ of both instruments (Sect.~\ref{sec:inv}).}
\label{fig:prof}
\end{figure*}
\begin{figure*}
\centering
\resizebox{\hsize}{!}{\includegraphics*{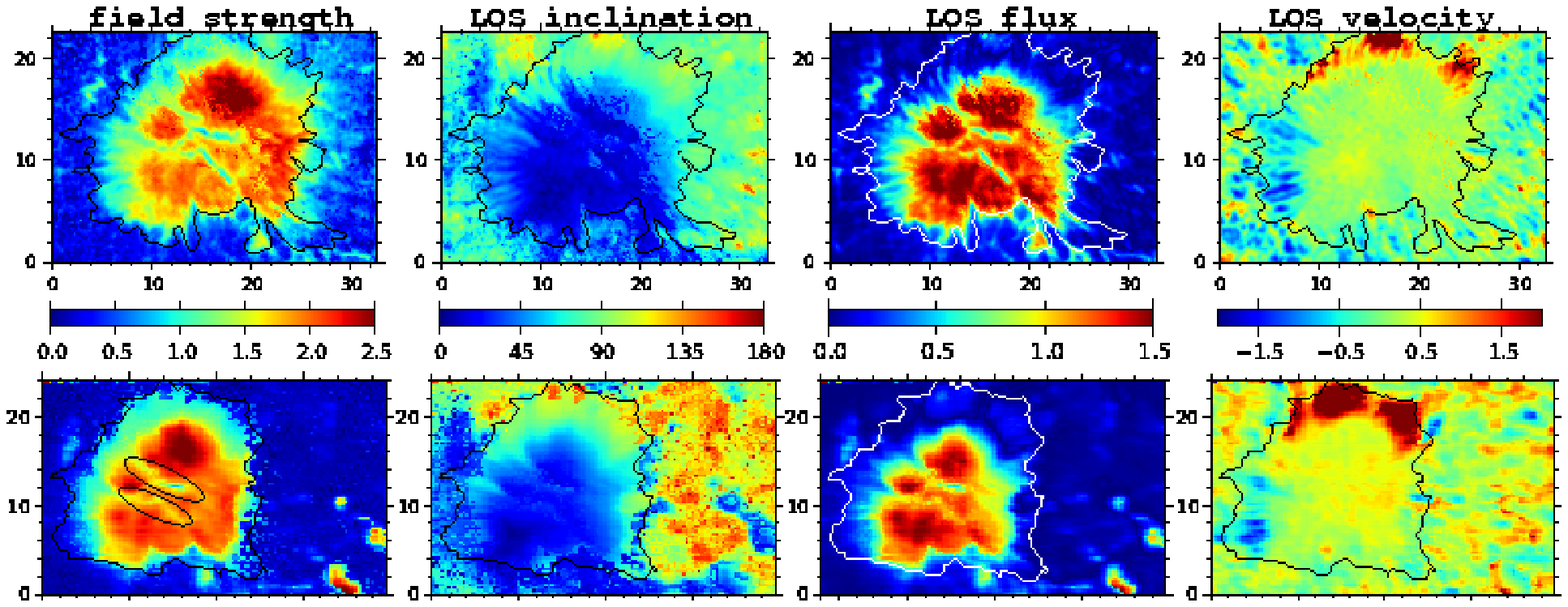}}
\caption{\emph{Left to right:} Maps of the magnetic field strength (kG), LOS inclination (deg), 
absolute value of the LOS flux ($10^{17}\,$Mx), and LOS velocity (km\,s$^{-1}$)  
for GFPI (\emph{top}) and TIP (\emph{bottom}) data.  
For TIP the magnetic flux is scaled by a factor of 1/9 to accommodate for its larger resolution elements. 
The ellipses in the field strength map of TIP mark the LBs. 
Other parameters are like Fig.\,\ref{fig:compare}.}
\label{fig:inv_comparison}
\end{figure*}

\subsection{Two-dimensional spectropolarimetric data}\label{sec:twodim}
The GFPI system scanned 31 spectral points along the Fe\,{\sc i} line at 617.3\,nm (Table\,\ref{tab:lines}) 
with a wavelength sampling of 1.48\,pm, recording the four Stokes parameters. 
The observations were taken in speckle mode, i.e. seven frames of short
exposure (20\,ms) per spectral position and polarimetric state, simultaneously taken with broad-band images.  
Using this method, we achieved a 1$\sigma$ noise level of about 8\,$\times$\,10$^{-3}\,I_{\mathrm c}$. 
The data have been filtered in wavelength to reduce noise and the residuals of the fluctuations of the 
spectrometer transmission by applying a filter with a $FWHM$ of 4.9\,nm in the Fourier domain \citep[see][Appendix A]{belloetal2009}.
An upgraded version of the ``G\"ottingen'' speckle code originally
developed by \cite{deboer_96} was applied to restore the broad-band data. This code takes 
into account the variations in the adaptive optics correction from the lock point 
in a similar way as explained in \cite{puschmann_sailer_06}.  
The spectropolarimetric data were then restored following \cite{KellerVdLuehe1992}, achieving
a spatial resolution of better than 0\farcs4 when seeing allowed. To minimize the
cadence to 56\,s, the readout time was reduced by exposing only half of the FOV of the GFPI detector.

We took care of the image shifts due to simultaneous scanning
steps of TIP by aligning the individual images. The reduced FOV typically amounts to
33$^{\prime\prime}$\,$\times$\,22$^{\prime\prime}$ with a spatial sampling of $0\farcs109$ per pixel.
The output of the GFPI narrow-band channel is a data cube containing spatial 
and spectropolarimetric information as a linear combination of four polarimetric 
states \citep{nazaret_kneer08}. To retrieve the Stokes parameters, 
the data are demodulated by applying the inverse Mueller matrix of the telescope-KAOS-GFPI
optical train. The telescope Mueller matrix is calculated following the VTT model by 
\cite{beck05b}. The Mueller matrix of the KAOS system plus the GFPI modulator
(polarimeter) is obtained from the calibration performed with the visible instrumental 
calibration unit as described in \cite{beck05a}. 
The estimated efficiencies on 2009 July 4 at $\sim$\,617.3\,nm are, for both
orthogonal\footnote{The GFPI polarimeter contains a polarizer beam-splitter that 
divides the incoming light into the {\em ordinary} and {\em extraordinary} (orthogonal) beams.
The advantages of this system are discussed in e.g. \cite{nazaret_kneer08}.}
polarimetric states, $\epsilon_\textrm{1}=(0.38, 0.43, 0.57)$ and
$\epsilon_\textrm{2}=(0.45, 0.44, 0.59)$, respectively. 
The data were corrected for crosstalk in the same manner as the TIP data.

\subsection{Line properties}\label{Sect:lines}
The \ion{Fe}{i} line at 1089.6\,nm has a line-core intensity of about 78\%$I_{\mathrm{c}}$ \citep{fts_atlas}. 
We measure a line-core intensity of 83\%\,$I_{\mathrm{c}}$ at a heliocentric angle of 28$^{\circ}$ with TIP (82\% at disk center). 
The small discrepancy can be explained by the presence of parasitic light (non-dispersed stray light) in the Echelle 
spectrograph  \citep[5\,\% in the visible,][]{beck_reza_11}
as well as by the existence of (dispersed) stray light \cite[10\,\%, ][]{beck_08}. 

The line depression response function for Stokes $I$ at line minimum peaks at $\log\tau\!=\!-1$ (150\,km) 
and the triplet Land\'e-factor amounts to 1.5, making \ion{Fe}{i} 1089.6\,nm a good choice to investigate the magnetic 
and velocity patterns in the lower photosphere (see Table\,\ref{tab:lines}). 
The Fe\,{\sc i} line at 617.3\,nm has a line minimum of 36\%$I_{\mathrm{c}}$ \citep[cf. BASS2000,][]{paletou_etal07}, 
and the GFPI measures a line minimum of 43\%$I_{\mathrm{c}}$ at disk center. 
The response function at line minimum peaks at $\log\tau\!=\!-1.5$ (230\,km),  
and the triplet Land\'e-factor amounts to 2.5. Therefore this line is well suited to studying magnetic fields at midphotosphere.
Figure\,\ref{fig:response} displays the response functions for the Stokes-$V$ profiles of the two lines\,\citep[cf.][]{cabrera_etal_05}. 
The values of the 2D response function is decoded in gray scale and displayed in $\log\tau$ versus $\Delta\lambda$. 
The response functions are calculated for a penumbra atmosphere with the magnetic and thermal stratifications 
from \cite{deltoro_iniesta_etal94} and \cite{luis_etal06}, respectively. While \ion{Fe}{i}\,1089.6\,nm shows a sizable 
response for $\log\tau> -2$, \ion{Fe}{i}\,617.3\,nm responds to changes in a slightly broader depth range $\log\tau> -2.5$.

\subsection{Time coverage} 
The sunspot evolution on July 4 was followed from 08:09 to 12:55\,UT with the GFPI, 07:58 to 13:11 with TIP, 
and 08:32 to 13:07 with the two imaging cameras. 

\section{Data analysis}\label{sec:analysis}
\begin{figure*}
\centering
\resizebox{\hsize}{!}{\includegraphics*{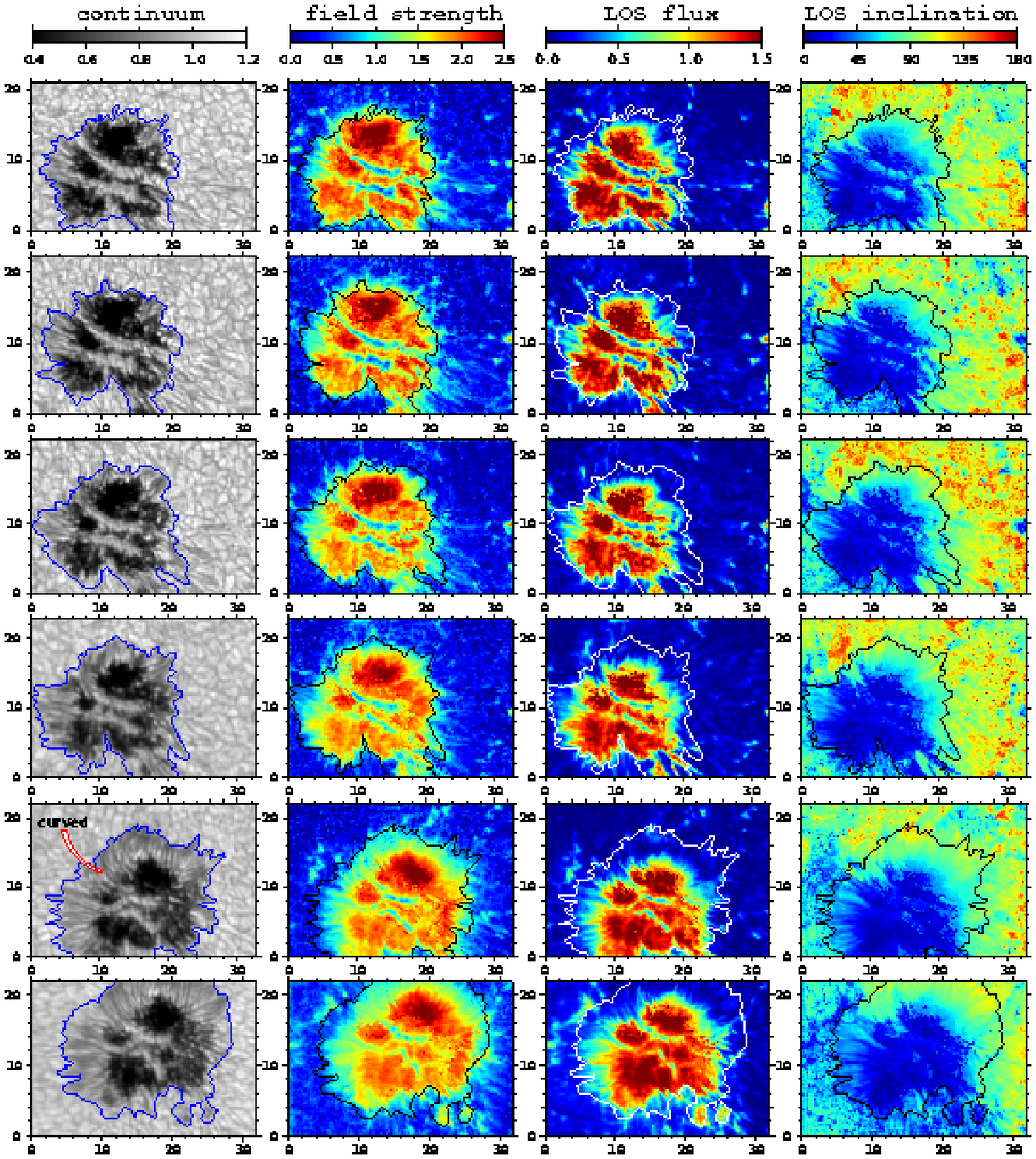}}
\caption{Temporal evolution of the physical parameters of the spot from GFPI data.   
From left to right we display maps of the continuum intensity (normalized to quiet Sun), 
magnetic field strength (kG), LOS magnetic flux ($10^{17}$\,Mx), and LOS 
inclination (deg). From top to bottom, the maps are taken 
at 08:40, 08:50, 09:28, 10:13, 11:51, and 12:38\,UT, respectively. 
The arrow in the continuum map at 11:51 points to a curved filament. 
Tick marks are in arcsec.}
\label{fig:table_maps}
\end{figure*}

\subsection{Typical data set} A snapshot of the NOAA\,11024 leading spot as seen in Stokes $(I, Q, U, V)$ 
at a late stage is shown in Fig.~\ref{fig:compare}.  The data were taken quasi-simultaneously 
with GFPI (upper row) and TIP (lower row). 
The TIP scanning direction is from the bottom to the top of the map. 
The contours that outline the spot were constructed manually from intensity maps. 
Although the spatial resolution of the TIP maps is lower than that of GFPI maps, 
and although the lines have a slightly different formation height, the maps compare very well. 
This comparison reassures us that the spectropolarimetric capabilities of the GFPI are suited to retrieve 
the physical parameters of the solar atmosphere.  Yet, one should keep in mind that the 
noise level in TIP data is lower by a factor of 8.

\begin{figure*}
\centering
\resizebox{\hsize}{!}{\includegraphics*{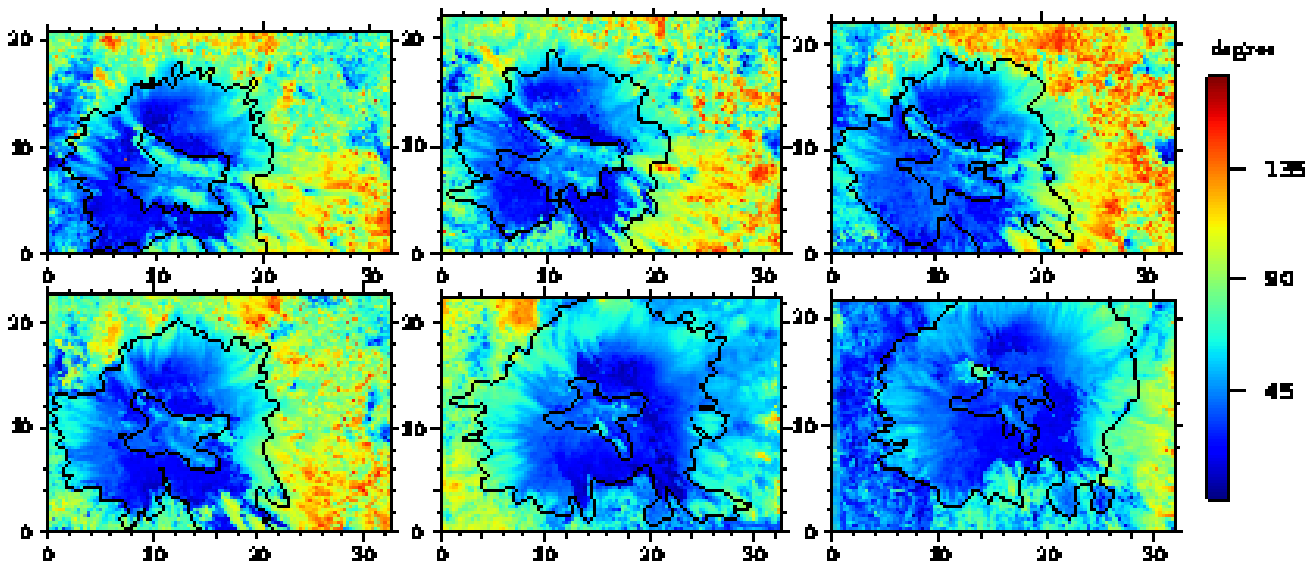}}
\caption{Temporal evolution, as in Fig.~\ref{fig:table_maps}, of the magnetic field inclination in the local reference frame.  
The maps correspond to 08:40, 08:50, 09:28, 10:13, 11:54, and 12:38\,UT. The inner contours outline LBs. 
The outer contours are the same as in Fig.~\ref{fig:table_maps}. }
\label{fig:local}
\end{figure*}

\subsection{Inversion of Stokes profiles}\label{sec:inv}
An inversion was performed using the SIR code~\citep{sir92, sir_luis} for TIP and GFPI data, separately. 
Some authors use two magnetic components in the inversion of 
sunspot penumbra~\citep[e.g.,][]{luis_etal_04, beck2010}. 
In order to keep the number of free parameters as small as possible, we did not use such a two-component setup. 
Instead, we used one magnetic component plus stray light, for the data of both instruments as in e.g.,~\cite{puschmann_etal_2010}. 
Such a one component and height-independent inversion setup has a slight tendency to underestimate the magnetic flux values in 
locations where strong gradients or opposite polarities \citep{reza_phd, morten_pdh} 
are present \citep{luis_rolf_etal_2003}.

The fraction of stray light was set to 15\,\%  for TIP \citep[][Sect. 5.4]{cbeck_pdh} and 12\,\% for 
GFPI \citep[][Sect. 3.2]{nbg_phd}, respectively, 
of an average quiet-Sun profile. 
\cite{cabrera_etal_07} and \cite{reza_etal_3} report similar values, but using different methods and 
definitions \citep[see also,][]{allendeprietoe+etal2004, beck_reza_damian_2011}.
In the single magnetic component, the magnetic field, the velocity, and the values for macro- 
and micro-turbulence were constant with height. 
Hence our inversion aims to estimate the average values of magnetic and velocity fields. We do not aim at modeling 
the asymmetries of the line profiles being produced by gradients along the LOS or by unresolved components.

Profiles of similar penumbral locations are compared in Fig.~\ref{fig:prof}. 
It demonstrate that the Fabry-P{\'e}rot spectropolarimeter 
is capable of recording complex profiles with only 31 wavelength points. 
As seen in Fig.~\ref{fig:prof}, the symmetric parts of the observed lines are well fitted, and we retrieve reliable
averages of the velocity and magnetic field parameters over the formation height of the lines.

Maps of the magnetic field strength, LOS inclination, magnetic flux along the LOS, 
and LOS velocity (as retrieved from the data sets of Fig.~\ref{fig:compare}) 
are compared in Fig.~\ref{fig:inv_comparison}. The upper row corresponds to 
GFPI and the lower one to TIP. The color codings for the maps of each quantity are identical for the two instruments 
except for the magnetic flux, which is the flux per pixel. Since the pixels of TIP correspond to an area on 
the solar surface, which is about a factor of 9 larger, the color coded values for TIP are rescaled correspondingly.

\subsection{Transformation to local reference frame (LRF)}
We convert the magnetic field inclination and azimuth from the LOS frame to the LRF 
using the known values of the position of the spot on the solar disk. 
Due to the intrinsic 180$^{\circ}$ ambiguity in azimuth \citep{metcalf94, lites_etal95}, we have two sets of solutions. 
Using the AZAM program \citep{elmore_etal92, tomczyk+etal1992, azam06}, we removed the ambiguity and reached the final results. 
%

\begin{figure}
\centering
\resizebox{\hsize}{!}{\includegraphics*{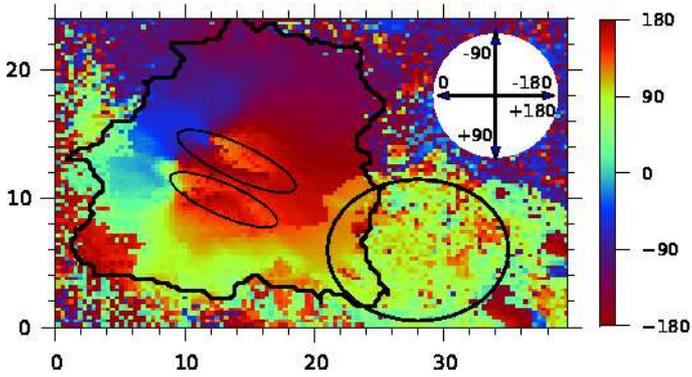}}
\caption{LRF azimuth map in the TIP data. The corresponding maps of other quantities are shown 
in the bottom panels of Figs.~\ref{fig:compare} and \ref{fig:inv_comparison}. Two thin ellipses mark 
the LBs. The thick ellipse shows 
the region where SFPs arrive. Zero degree azimuth corresponds to the $(-x)$ direction, and $\pm\,180^\circ$ 
correspond to the $(+x)$ direction (see arrows in the white circle). The abscissa is in arcsec.}
\label{fig:azimuth}
\end{figure}

\section{Magnetic evolution}\label{sec:magnetic}
A sequence of six snapshots tracking the formation of the penumbra is shown in Fig.~\ref{fig:table_maps}. 
The contours were drawn manually as the intensity boundary of the spot. From left to right we display maps of intensity, 
field strength (kG), magnetic flux (LOS, $10^{17}$\,Mx), and inclination (LOS, degrees). From top to bottom time evolves 
from 08:40, 08:50, 09:28, 10:13, 11:51, and 12:58\,UT. A legend for the color coding is displayed at the top of each column. 
An additional set of maps at 11:54\,UT is shown in the top row of Fig.~\ref{fig:inv_comparison}.

\begin{figure}
\centering
\resizebox{\hsize}{!}{\includegraphics*{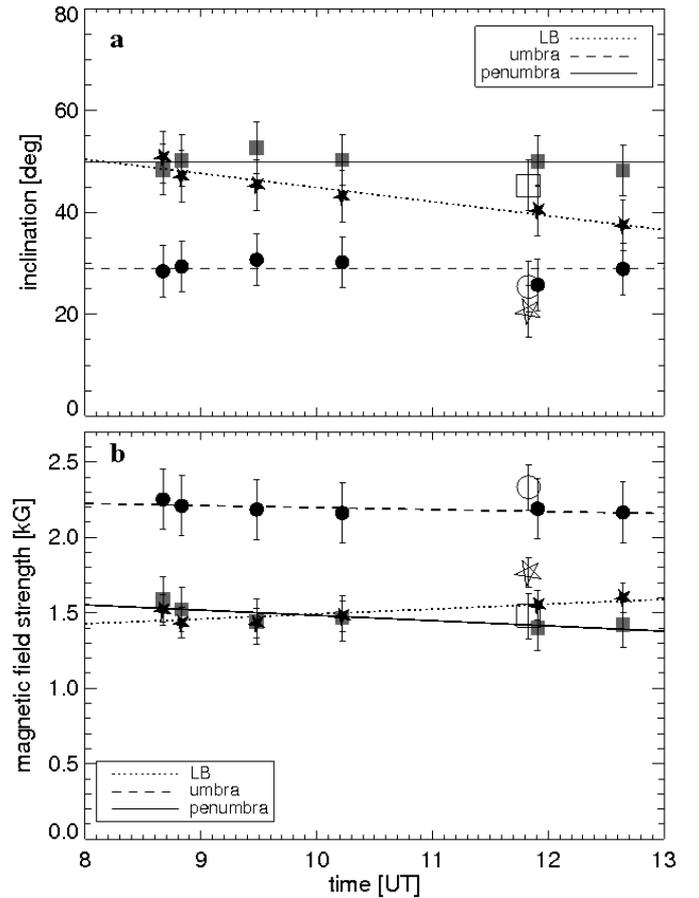}}
\caption{Variation in the LRF inclination angle (top) and  magnetic field strength (bottom) 
in the umbra\,(dashed, circle), penumbra (solid, square), and 
LBs (dotted, asterisk). The corresponding TIP values are shown as empty symbols. 
The time steps are the ones shown in Fig.~\ref{fig:local}.}
\label{fig:inclin}
\end{figure}

Figure~\ref{fig:local} shows the maps of the LRF field inclination, $\alpha$, for the selected snapshots. 
The azimuth maps display mostly radial field (cf. the TIP azimuth map in Fig.~\ref{fig:azimuth}). 
Two ellipses that mark the light bridges (LBs) are also shown in the 
field strength map of TIP (bottom left panel, Fig.~\ref{fig:inv_comparison}). 
The manual contours of the outer intensity boundary in Fig.~\ref{fig:local}  
are identical with those of Fig.~\ref{fig:table_maps}. The inner contours mark the LB boundaries. In areas 
outside the spot where the magnetic field strength is low (cf. Fig.~\ref{fig:table_maps}), the amplitudes of  
$Q$ and $U$ are too small to provide reliable estimates for the inclination \citep[see ][]{borrero_etal_2011}.

\begin{figure*}
\centering
\resizebox{\hsize}{!}{\includegraphics*{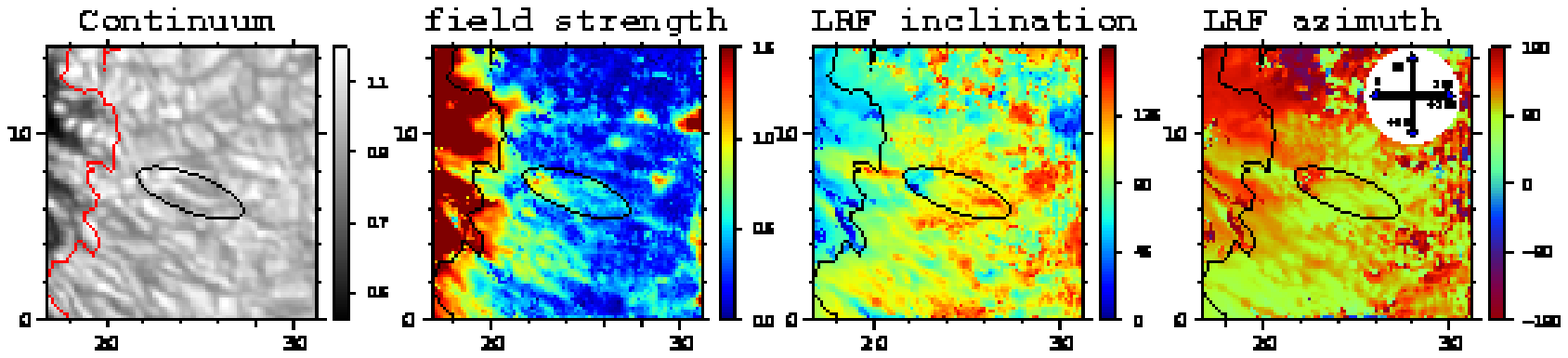}}
\caption{Elongated granule and the corresponding bipole (marked by ellipse) in GFPI data. 
{\emph{Left to right:}} continuum intensity (normalized to quiet Sun), 
magnetic field strength between zero and 1.5\,kG, inclination, and azimuth (deg) in the local reference frame. 
The coordinates of the selected area shown here correspond to the middle-top panel of 
Fig.~\ref{fig:local}.}
\label{fig:bipole}
\end{figure*}

%
Average values for the magnetic field inclination of umbra, penumbra, and LB 
are plotted in Fig.~\ref{fig:inclin}a versus evolution time. The evolution of the average magnetic field strength 
is plotted in Fig.~\ref{fig:inclin}b. 
The average inclination of the magnetic field retrieved from the TIP data {is less than 10$^{\circ}$ smaller 
(i.e., more vertical)} than that of the GFPI data, both in the umbra and penumbra. Assuming that the magnetic field 
fans out with height,
we ascribe this difference to the lower formation height of the TIP line (cf. Fig.~\ref{fig:response}).
The relative large difference in the inclination of LBs between TIP and GFPI is due to the lower spatial resolution 
and a larger pixel size in TIP, leading to a larger contamination by umbral light. This is also seen to some extent in the magnetic 
field strength of the LBs in TIP data: it has a higher value, meaning that it was contaminated by umbral pixels.

\subsection{Intensity and magnetic spot boundaries}
The contours drawn in the maps of Figs.~\ref{fig:table_maps} and \ref{fig:local} are based on continuum intensity. 
As the maps of magnetic field strength demonstrate, kilo Gauss fields in many cases extend beyond the intensity contours. 
They are stronger than a typical canopy field \citep[e.g.,][]{reza_etal_1}. 
Towards the lower right, i.e., towards the emergence site, all evolution snapshots show strong magnetic fields of more 
than 1\,kG. 
These fields come from emerging bipoles and, as discussed in the next subsection, partially merge with the spot.

Large extensions of magnetic field beyond the intensity contours are also seen in the upper left hand side of the spot in the 
first three snapshots of Fig.~\ref{fig:table_maps}, e.g., in the first map at ($x$\,$=$\,$6$,\,$y$\,$=$\,$15$). 
As the penumbra develops on this side of the spot, these extensions become smaller. After the penumbra has formed, 
the intensity contours coincide well with the magnetic boundaries of the spot. That means that the magnetic flux 
that was initially outside the spot was incorporated into the spot during the penumbra formation process. 
The amount of flux, however, is only very small in the flux balance, since the field inclination 
is close to horizontal in these areas.

\subsection{Supply of magnetic flux from moving SFPs}\label{sec:elongated}
In Papers\,1 and 2, we reported on the properties {and processes} at the flux emergence site close to the observed 
spot: there, elongated granules are the visible signature of flux emergence occurring on granular scales in 
ARs. They can be traced in maps of magnetic field strength and 
inclination. They have a typical size of 2--5\,Mm\,$\times$\,0.35\,Mm. Sometimes dark lanes with similar dimensions 
are observed \citep[Paper\,1, see also][]{brants_steen_85}. Their magnetic counterparts are bipolar 
features with the poles being cospatial with the ends of the elongated granules (Fig.~\ref{fig:bipole}). As they evolve, these bipoles 
dissociate such that their poles separate and migrate towards their proper AR polarity.  
An example of such an evolution was shown in Fig.\,2 of Paper\,2, marked with an ellipse in Fig.~\ref{fig:bipole}. 
The LRF inclination and azimuth maps of that example, as well as the field strength and continuum intensity 
are seen in this figure. 
The field azimuth in between the opposite polarities is along the axis of the elongated 
granule (marked with the ellipse). 
The two ends of the elongated granule are also cospatial with opposite polarities as seen in the inclination panel. 
In the field strength map, fields of some 1.1\,kG are seen in the polarity closer to the spot, while in the other 
pole, the field strength amounts only to about 700\,G. 
These small flux patches (SFPs), which are close to the forming spot and which have the spot polarity, finally merged with it. 
As seen in Figs.~\ref{fig:table_maps} and \ref{fig:local}, e.g. (x=20,\,y=8) or the inclination panel of Fig.~\ref{fig:bipole}, 
SFPs of both polarities exist close to the spot in the initial stages.

Using a sample of some 100 SFPs from the snapshots shown in Fig.~\ref{fig:table_maps},  we measured a 
magnetic flux of 2--3\,$\times\,10^{18}$\,Mx for a typical SFP, comparable to the flux values reported by \cite{wang_zirin_92}. 
We also detect a number of SFPs that are larger than $10^{19}$\,Mx, i.e., comparable to little pores.
For comparison, the magnetic flux measured in a granule located {some 20\arcsec\  away from} the spot 
(in a direction perpendicular to the AR axis) is an order of magnitude smaller.
And assuming that the average (vertical) field strength in the quiet Sun amounts to 11\,G \citep{lites_etal_07b}, 
a granule with a diameter of $1\arcsec$ amounts to a flux of  $4\times 10^{16}$\,Mx.

The magnetic field strength at the ends of elongated granules (which are in the vicinity of the spot) 
is typically about 1\,kG and the LRF inclination is about 70$^{\circ}$.  
Averaging over a sample of 100 pixels in the central area of elongated granules, the 
field strength is $B\!\approx\!400\,$G ($RMS\!=$\,200\,G), and the field inclination 
is horizontal within 20$^{\circ}$. 
Similar values in magnetic field strengths have been found by \cite{kubo_etal03}.

As shown in the right hand panel of Fig.~\ref{fig:bipole}, the SFPs show a very interesting behavior when they approach the spot.  
They have the same azimuth as the spot, 
meaning that their magnetic field is aligned with field of the spot (Fig.~\ref{fig:azimuth})
\footnote{The azimuth maps in Figs~\ref{fig:bipole} and \ref{fig:azimuth} have identical color tables.}.

\begin{figure}
\centering
\resizebox{\hsize}{!}{\includegraphics*{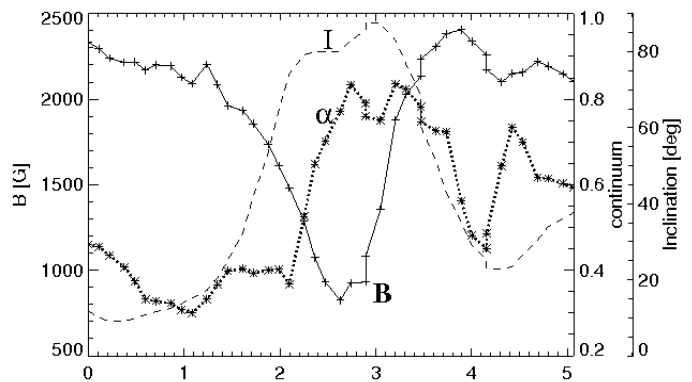}}
\caption{Variation in the continuum intensity (\emph{dashed}), magnetic field strength (\emph{solid}), 
and inclination (dotted) across the upper left LB in the second top panel of Fig.~\ref{fig:table_maps}. 
The abscissa is distance in arcsec, and the cut starts at the upper right of the LB.}
\label{fig:lightbridge}
\end{figure}

\subsection{Light bridges}\label{sec:lbs}
Light bridges are elongated areas in the umbra with enhanced intensity \citep{sobotka_etal_93}. 
They have a fine structure consisting of granular-like segments (see Paper 1). 
However, the top-left end of the LB (Fig.~\ref{fig:table_maps}) is not segmented, 
similar to the findings of \cite{lites_etal_04}. 
LBs in the umbra typically have a width of $<$\,2\,Mm and show up with enhanced intensity. 
They correspond to locations in which the magnetic field strength 
is significantly weaker than in the surrounding umbra, as one can see in the field strength maps of 
Fig.~\ref{fig:table_maps} (second column).
They also show lower magnetic flux and more inclined fields (third and fourth columns, respectively).  
As seen in Fig.~\ref{fig:table_maps} and better in G-band and Ca images (Paper\,1), LBs have a substructure: 
convection-like cells, and dark lanes. The dark lanes are not exactly located at the center of the LBs, 
but always towards the solar limb, being consistent with the idea that LBs are symmetric: 
Since the $(\tau\!=\!1)$-level is elevated relative to the surrounding umbra~\citep{lites_etal_04}, the center side of the 
LB appears larger than the limb side, and the central lane is projected towards the solar limb.

Figure~\ref{fig:lightbridge} shows the continuum intensity, magnetic field strength, 
and LRF inclination angle along a cut across 
the upper left LB (marked in the continuum map of the second top panel in Fig.~\ref{fig:table_maps}). 
The cut starts on the upper right (limb side) and crosses the LB towards the lower left (center side).
The LB brightness enhancement reaches up to intensity values of average granulation. 
The dark lane signature can be seen at the top of the intensity curve as reported earlier 
\citep{sobotka_etal_94,berger_berdyugina_2003,giordano_etal_2008}. 
The field lines in the dark lane are more vertical compared to the bright segments (Fig.~\ref{fig:lightbridge}). 
Such dark lanes show a higher contrast in Ca\,{\sc ii}\,K images (see Paper 1, Fig.\,3). 
In the LB, the magnetic field strength drops below 1\,kG with a minimum value that 
spatially coincides with the LB dark lane.
In the LRF inclination maps (Fig.~\ref{fig:local}), the LBs are outlined with contours (manual drawing based on intensity maps).
The LB field lines are more inclined relative to the surrounding umbral field, reaching inclinations 
up to 50$^\circ$ relative to the vertical. The azimuth map at about 11:50\,UT is shown in Fig.~\ref{fig:azimuth}. 
In this map, two thin ellipses mark the LBs. The field lines in the LBs are roughly 
aligned with the orientation of the LBs as previously reported by \cite{louis_etal_2009}.

During the sunspot evolution, the average LRF inclination in the LBs gradually decreases 
from $\sim$\,50$^{\circ}$ to $\sim$\,37$^{\circ}$; i.e., the field gradually becomes more vertical 
(Fig.~\ref{fig:inclin}a). The change in the average magnetic field strength of the LBs is shown in
Fig.~\ref{fig:inclin}b, where it increases from 1.4 to 1.6\,kG. As the magnetic properties of the LB approach those 
of the surrounding umbra, the LB area shrinks in the intensity images.

\subsection{Penumbra}\label{sec:formation} 
\begin{figure}
\centering
\resizebox{\hsize}{!}{\includegraphics*{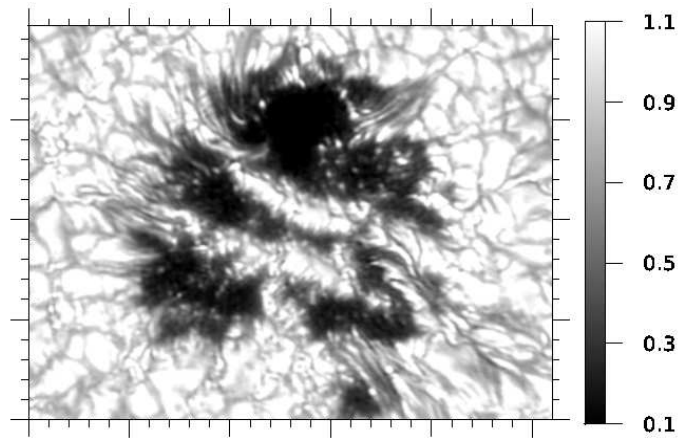}}
\caption{G-band snapshot of the forming sunspot at 08:33\,UT. The intensity has been clipped to lie between 
0.1 and 1.1 (1.0 is the average quiet Sun). Each tick mark is one arc second.}
\label{fig:saturated}
\end{figure}

The development of the penumbra can be followed from the snapshots in Fig.~\ref{fig:table_maps} 
and from the G-band movie published together with Paper 1. In Paper 1 we stated that 
transient penumbral filaments form all around the protospot, but the locations in which stable 
filaments form is limited to a few sections.
At the earliest stage of our observations, all penumbral sections are found next to the ends of LBs 
(see, e.g., $x\!=5$,\,$y\!=$10). 
The penumbral section around ($x\!=$18,\,$y\!=$14) is not next to one of the major LBs, 
but as one can clearly see in Fig.~\ref{fig:saturated} it is located next to an area of diffuse umbral dots, 
which ``connect'' the penumbral filaments to the major LB.

These sections grow outwards and all around the penumbra except for the lower right hand side, where only a 
few transient filaments form. There it seems that the emerging flux and the approaching SFPs inhibit the 
formation of stable filaments. This indicates that, for the penumbra to form, a quiet surrounding in which 
the penumbra can expand and evolve is necessary.

While the penumbra develops, the average penumbral magnetic field strength keeps constant around 1500\,G 
(Fig.~\ref{fig:inclin}b), showing an insignificant decrease ($\delta\!B\sim100$\,G). 
The average LRF inclination also stays roughly 
constant at $\sim$\,50$^{\circ}$ ($\delta\theta\sim5^{\circ}$, see Fig.~\ref{fig:inclin}a). 
These are typical values for the penumbrae 
\citep{lites_skumanich_90, keppen_martinez_96, mathew+etal2003,luis_04, beck_08, tritschler09}. 
On the border between the penumbra and quiet Sun, the average field strength decreases from 1000\,G to 800\,G.

The filamentary structure of the penumbra can be seen in both intensity and magnetic maps. 
The magnetic field inclination, both in the LOS and LRF frames (Figs.~\ref{fig:table_maps} and \ref{fig:local}), 
shows clear signs of spine and interspine structures \citep{lites_elmore_etal93,cuberes_etal_05}. 
The peak-to-peak difference in the LRF inclination between spines and interspines is about 10$^{\circ}$--20$^{\circ}$.  
This variation is smaller than the corresponding values for a mature spot~\citep[e.g.,][]{kubo_etal_07}.  
However, such a peak-to-peak difference is a function of radial distance from the spot center~\citep{westendorp+etal2001}. 

In the existing penumbral sections for the snapshot at 08:40\,UT at around ($x$\,$=$\,4, $y$\,$=$\,10), the 
LRF inclination increases radially 
from $40^{\circ}$ (inner penumbra) to  $65^{\circ}$ (outer penumbra)  and at ($x$\,$=$\,16, $y$\,$=$\,12) 
from $40^{\circ}$ to $70^{\circ}$, and keeps nearly constant with time. 
In contrast, the inclination increases with time as the penumbra advances 
in the new forming section, e.g. around ($x$\,$=$\,12, $y$\,$=$\,18) in the snapshot at 08:40\,UT 

We aligned the larger part of umbra in the first (08:40\,UT) and last (12:38\,UT) stages (Fig.~\ref{fig:table_maps}).
While the field lines become more inclined with the developing penumbra, we find that
the magnetic neutral line keeps the same distance to the umbra.

\subsection{Curved penumbral filaments}\label{sec:curved}
At late stages, the upper penumbra 
shows filaments that are not radial but curved (Fig.~\ref{fig:table_maps}). 
The curvature, as seen in the continuum and G-band images, amounts to more than 45$^{\circ}$. 
Beside that, as seen in the bottom middle panel of Fig.~\ref{fig:local} ($x\!=\,16$,\,$y\!=\,22$), the 
field lines change their {\it polarity} in that location. From other snapshots not shown here, we find that 
there are kilo Gauss fields outside the spot and close to these curved area. 
Parts of this opposite-polarity area coincide with an enhanced brightening in the cotemporal \ion{Ca}{\sc ii}\,K image. 
Reported amounts of curvature in the penumbral filaments are usually 
less \citep{lites_skumanich_90, keppen_martinez_96, westendorp+etal2001}. 
\cite{gurman_house81}, however, reported a twist angle of 35$^{\circ}$. 
From the maps of the same spot in the later days (Paper 1, Fig.\,2), we know 
that the curved filaments become essentially radial in subsequent days. 

\subsection{Umbra}
Figure~\ref{fig:inclin}b shows the evolution of the umbral magnetic field strength, which stays constant at 2.2\,kG during 
the evolution. The maximum field strength -- determined as the average of 100 umbral pixels that show the strongest 
field strength values -- also stays constant at  $\sim$\,2.7\,$\pm$\,0.2\,kG (not shown). The average of the umbral 
inclinations in the LRF remains constant at some 28$^{\circ}$. 
However, the inclination in the umbra is not uniform, especially in the lower part. 
There, at the border between the umbra and quiet Sun the inclination is less than 10$^{\circ}$ and  increases 
gradually from there toward the spot center.

Another important finding is that the small umbra in the left hand section shrinks in size, while its 
mean inclination increases from 30$^{\circ}$ to 40$^{\circ}$. 
In the meantime, a new umbral core develops towards the emergence region, where no penumbra forms. 
Thus, the net umbral area remains constant during the spot's evolution (Paper\,1).

Figure~\ref{fig:flux} shows the variation in the total magnetic flux with time, both in the LOS and LRF frames. 
As expected, the values are higher in the LRF. 
The magnetic flux of the penumbra is larger than the flux in the 
umbra (or umbra + LBs), already from early stages. 
That the majority of the magnetic flux is in the penumbra is a universal property of sunspots 
\citep{hu_schmidt_91, solanki_schmidt92, balthasar_collados_05}, 
and suggests that the penumbra is not flat, but extends into deep subphotospheric layers.

\subsection{Flux budget}
Using the G-band and \ion{Ca}{ii}\,K time series (Paper 1), we have reported that the umbral area stays 
constant (100 arcsec$^2$) while the total sunspot area increases from 230 to 360 arcsec$^2$ (an increase of 56\,\%). That means 
the penumbral area increased from 130 to 260 arcsec$^2$, corresponding to an increase of 100\,\%. 
This increase in area reflects the increase in magnetic flux. 
Here, we determine the increase in the magnetic flux, 
using the magnetic field strength, $B$, and the inclination angle in the LOS frame, $\gamma$, or 
in the LRF, $\alpha$ (Sect.~\ref{sec:analysis}):
\[ \Phi = A\, B\, \cos\alpha \]
where $A$ is the deprojected area of each pixel.

\begin{figure}
\centering
\resizebox{\hsize}{!}{\includegraphics*{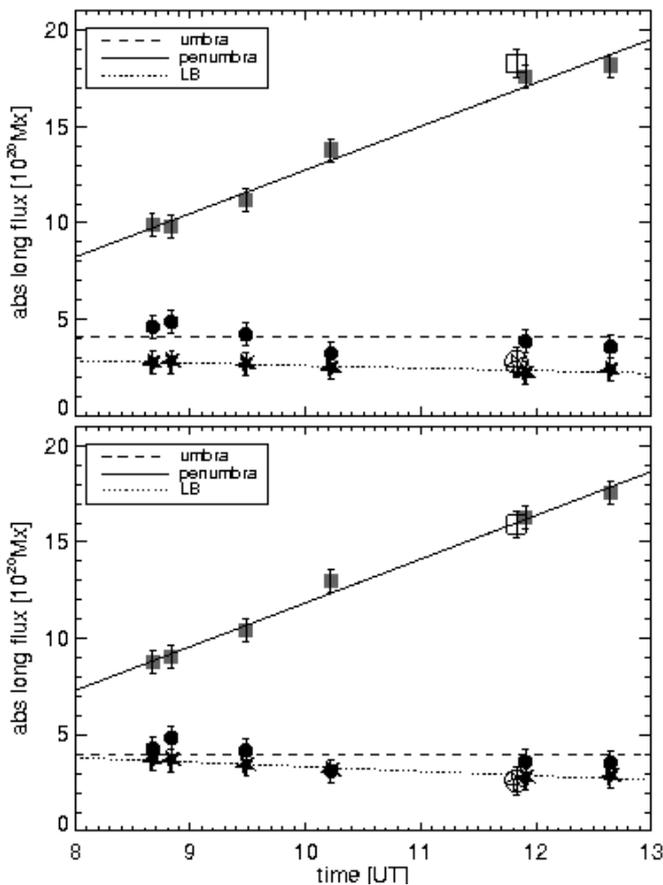}}
\caption{Similar to Fig.\ref{fig:inclin} but for the magnetic flux. \emph{Top}: LRF, \emph{bottom:} LOS frame.}
\label{fig:flux}
\end{figure}
The umbral area was defined as the area of dark pixels with a continuum intensity lower than 0.5\,$I_{\mathrm{c}}$. 
For the penumbra and LBs, we used   manual contours. The penumbral contours are shown in Fig.~\ref{fig:table_maps}, while 
both penumbra and LBs contours are shown in Fig.~\ref{fig:local}. 
As mentioned before, the average field strength of the umbra is constant. Considering that the umbral area 
keeps constant (Paper 1, Fig.\,4), we find that the total {\em umbral} flux is constant within our errors. 
There is a slight tendency of a decreasing umbral flux as is the case for the area (Paper 1).

The umbrae show strong dynamics during the sunspot evolution. For instance, parts of the LBs dissolve into 
umbral dots and diffused emission, while parts of the umbra shrink at expenses of the penumbral formation 
(left umbrae) and other umbral subareas change in size. 
However, the overall result is that the umbral area, field strength, and flux stay constant.

As mentioned above, the sunspot (penumbral) area increased by about 56\,\% (100\,\%). 
We find that the penumbral flux increases from 
9.7\,$\times\,10^{20}$ to 18.2\,$\times\,10^{20}$\,Mx (88\,\%)\footnote{
The last snapshot does not fully cover the spot (Fig.~\ref{fig:table_maps}), 
so the true amount of the flux is a few percent more.}.
The void symbols in Fig.~\ref{fig:flux} represent the corresponding values for the TIP  
map shown in Fig.~\ref{fig:compare}\footnote{The spatial sampling in the TIP data is larger than GFPI, 
so there is more uncertainty in the manual contours, leading to larger uncertainties in 
the total magnetic flux.}.

\subsection{Summary of results}\label{sec:results}
We now summarize the magnetic properties of the forming spot presented in the previous sections.
\begin{enumerate}
 \item A strong ($\approx$\,1\,kG) inclined (45$^\circ$-\,60$^\circ$) magnetic field component is present outside 
the visible spot boundaries previous to the penumbra formation. 

 \item The maximum ($\sim$\,2.7\,kG) and the spatially averaged (2.2\,kG) magnetic field strength of 
the umbra, stays constant during the penumbra formation.
In addition,  the penumbral field strength slightly decreases around 1.5\,kG, and in the LB it increases from 
1.4\,kG to 1.6\,kG (Fig.~\ref{fig:inclin}b). 

 \item The average LRF magnetic field inclination in the umbra  (28$^{\circ}$) and penumbra (50$^{\circ}$) 
do not change  during the spot formation, while in the LBs it decreases from 50$^{\circ}$ to 37$^{\circ}$; i.e., the 
field lines in the LBs become more vertical (Fig.~\ref{fig:inclin}a).

 \item The total magnetic flux of the spot increased by 40\,\%, from 17.4\,$\times\,10^{20}$ to 24.2\,$\times\,10^{20}$\,Mx. 
This is consistent with the earlier report of a 56\,\% increase in the spot area (Paper 1). 
The increase in the penumbral area is of about 100\,\%, 
corresponding to an increase of the magnetic flux of 88\,\% (Fig.\ref{fig:flux}). 
While the flux in the umbra keeps constant, the magnetic flux of the LBs decreased from 2.8\,$\times\,10^{20}$ 
to 2.4\,$\times\,10^{20}$\,Mx (its area shrinks).

 \item Averaging over some 100 SFPs, we find that the mean magnetic flux of individual 
SFPs amounts to 2--3\,$\times\,10^{18}$\,Mx. The typical value of the inclination of the field lines in SFPs 
close to the spot is about 70$^{\circ}$.

 \item The average rate of flux accumulation in the spot is 4.2\,$\times\,10^{16}$\,Mx\,s$^{-1}$, which 
is equal to the merge of one SFP per minute to the spot. 

 \item The new magnetic field lines reach the spot boundary with the same azimuth angle as the spot azimuth; i.e., 
both SFP and spot field lines are coaligned.

 \item The new flux arrives in the side of the spot facing the AR opposite polarity, while the penumbra develops on 
the other side of the spot. 
During this process, the position of the magnetic neutral line relative to the umbra does not change. 

 \item The first penumbral sections preferably form close to the ends of LBs. 
As the penumbral sections form, the inclination is increasing.

\end{enumerate}

\section{Discussion}\label{sec:discussion}
In this section we attempt to construct a consistent scenario of the spot formation process by assembling our results 
with previous findings. After sorting out the involved timescales (Sect.~\ref{sec:timescale}), we discuss the properties 
of the emergence site and the transport of flux to the spot (Sects.~\ref{sec:egran} and \ref{sec:transport}). 
We ascribe a crucial role to LBs, as remnants of trapped granules between merging pores.  
They ignite the formation of the penumbra 
and allow flux to be transported from one side of the spot to the other (Sects.~\ref{sec:remnant}, \ref{sec:ignitor}, 
and \ref{sec:channel}). In the last section we elaborate on the necessities to form a penumbra and discuss the 
properties of the developing penumbra.

\subsection{Timescales in a sunspot}\label{sec:timescale}
Sunspots live for days to weeks~\citep{bray_loughhead_64}. 
Next to the long lifetime ($t_{\rm l}\!\sim$\,a week), there is a formation timescale. 
This is the time required to assemble a spot out of disorganized emerged flux. 
We estimate the formation time to be about $t_{\rm f}\!\sim$\,0.5\,day. 
The dynamical timescale is the time that the spot requires to rearrange the equilibrium configuration of the
field lines in a smooth transition between two equilibrium states. 
The dynamical time corresponds to the Alfv\'en travel time across the spot. 
The Alfv\'en and sound travel speeds (at $\tau\!=\!1$) are 
about  $v_{\mathrm A}\!\sim$\,8\,km\,s$^{-1}$, $c_{\mathrm s}\!\sim$\,6\,km\,s$^{-1}$, respectively \citep{rolf_phd}. 
The travel time in our sunspot with d\,$\sim\,$15\,Mm is $t_{\rm d}\!\sim$1\,h.
Individual penumbral filaments also form on a dynamical timescale. 
Smaller timescales, $t_{\rm s}\lesssim$\,5\,min, have been observed 
by \citet[][their Fig. 5]{sobotka_etal97} and \cite{ortiz_etal10} in umbral dots and LBs.  In summary the 
timescales, $$t_{\rm l}\,\gg\,t_{\rm f}\,\gg\,t_{\rm d}\,\gg\,t_{\rm s}$$ correspond to different physical phenomena: 
the decay process, the integration of the new flux, the penumbral formation, and probably some type of convective 
timescale, respectively.

\subsection{Elongated granules}\label{sec:egran}
The AR flux emergence site close to the leading spot is characterized by elongated granules and extended dark lanes,  
together with micropores (circular patches of reduced intensity) 
as seen in intensity \citep[see Paper\,1 and e.g.,][]{gomory_etal_2010}. 
The properties of the elongated granules found in our data are similar to those observed 
by \cite{brants_steen_85} in a forming AR, 
by \cite{strous_zwaan99} in a developing AR, and 
by \cite{otsuji_etal_07} in the moat of a fully fledged sunspot. 
They are also found in numerical simulations of flux emergence in 
ephemeral \citep{cheung_etal07,cheung_etal08} and active \citep{cheung_etal10} regions.

The magnetic field strength in the emergence site is weaker than in the vicinity of the spot (see Sect.~\ref{sec:elongated}). 
Some magnetic field intensification must then work on the (hG) emerged flux to facilitate the kilo Gauss field 
strength observed at the ends of elongated granules~\citep[e.g.,][]{weiss66, spruit_79,grossman_etal_98}. 
These mechanisms have some observational support~\citep{luis_etal_01, nagata_etal_08, danilovic_etal10}. 

In our observations, elongated granules appear only in a limited cone close to the spot toward the opposite 
AR polarity ($\sim$ a quadrant). In contrast, the simulation by \cite{cheung_etal10} shows elongated granules 
almost isotropically around the spot. This might be due to the limited separation of the two polarities 
in the simulation domain and possibly an effect of periodic  boundary conditions. 

Elongated granules carry significant amounts of magnetic flux, but have similar continuum intensities compared to the 
quiet Sun \citep[cf.][who call them radiatively undisturbed points]{leka_sku98}. On the opposite ends of some elongated 
granules the magnetic field has opposite polarities. 
To understand the overall field topology, the orientations of these bipoles is crucial. In a simple model, 
the SFPs having the same polarity as the spot finally merge with it. 
However, observations demonstrate that the topology is typically 
more complex \citep[e.g.,][]{bernasconi_etal_2002,  pariat_etal04}. 
As explanations, undulations of field lines are discussed as due to their interactions with the granular convection 
and resistive effects like magnetic reconnection \citep[e.g.,][]{strous_zwaan99, cheung_etal08}. 
We find elongated granules with field lines of opposite polarity at either end, but our field of view and 
the temporal resolution are insufficient to reconstruct the evolution of the field topology of our emergence site.

\subsection{Flux transport in the active region}\label{sec:transport}

On July 4, at 08:30 UT the protospot had a flux of 17.4\,$\times\,10^{20}$\,Mx. In 4.5\,h, it transformed into a mature spot 
with a major penumbra. Two questions arise. (1) How did the initial protospot form? (2) How did the penumbra grow in size 
and in magnetic flux?

{(1) Merging pores:} 
From SoHO/MDI data it is seen that two pores appeared next to each other at 20:47 UT on July 3 \citep{rolf_etal_10c}. 
These pores increase in size, merge, and form the protospot, in accordance to the proposal of \cite{zwaan92}. 

{(2) SFPs supply flux:} In the 4.5 h of the VTT observation no further pore merges with the spot. 
Although pores are observed to merge later again, the flux increase by $\sim 7\,\times\,10^{20}$\,Mx between 08:30 UT 
and 13:00 UT cannot be due to merging pores. The tiny pore below the spot is included in our contours, i.e., 
merged before 08:30. The increase in the spot's flux is comparable to the flux of a few pores. 
Since we see examples of SFPs that merge with the spot, we propose that all the additional flux was carried to the spot by SFPs.
Although the coalescence of pores \citep{zwaan92} is therefore an important step in the formation 
of protospots (initial phase), a majority of the extra flux required to form the penumbra 
is provided via SFPs. In our case, an increase rate of 4.2\,$\times\,10^{16}$\,Mx\,s$^{-1}$ 
needs to be supplied by merging SFPs; i.e., one SFP per minute is needed.

We can also compute a value for the flux increase rate for forming the protospot. The flux emergence started latest on July 3, 
at 20:47\,UT \citep[see][using MDI data]{rolf_etal_10c}. Using the known value of the flux of the 
protospot, we estimate that the {\it initial} flux rate amounts to 3.9\,$\times\,$10$^{16}$Mx\,s$^{-1}$. 
These two rate values agree with each other and with previous observations~\citep[e.g.,][]{kubo_etal03} 
and comparable to the  values of \citet[][their Fig.\,10]{cheung_etal10} from an intermediate phase in the simulation. 
For a pore that develops a partial penumbra \citet{leka_sku98} give a lower rate of 1\,$\times\,$10$^{16}$Mx\,s$^{-1}$.

From these considerations, we presume that pores also grow by merging SFPs. In that sense, elongated granules, 
emerging bipoles, and the corresponding SFPs are the building blocks of large-scale magnetic structures in the photosphere. 
SFPs build up pores, and then spots are formed out of pores and SFPs.

\subsection{Light bridges as remnants of spot formation} \label{sec:remnant}

During the process of merging pores, a LB forms as 
the granular area in between the pores that are ``trapped''. Using SoHO/MDI data \citep{rolf_etal_10c}, 
we find that at least two pores of the same polarity emerged close to each other on July 3, 
and merged to form the proto-spot on July 4. In that sense, the LBs that we observe on July 4 are remnants 
of the formation process of the protospot. A light bridge that remains after a pore has merged can 
also be seen on July 5 \citep[see Fig.\,2 in Paper 1, and also][]{roupe_luis_etal_2010}. 
During the later evolutionary stages other types of LBs exist \citep[e.g.,][]{katsukawa_etal_2007}. 
\citet{garcia_87} proposes that LBs that form in the decaying phase are related 
to the ones that faded away in the forming phase. 

As seen in Sect.~\ref{sec:lbs}, LBs have a weaker and more inclined magnetic field than the umbra.  
Such a reduced magnetic field strength agree with \citet{rueedi_etal_95} and \citet{leka1997}. 
We estimate that the horizontal gradients of the magnetic field strength 
in the LBs is about 1\,G\,km$^{-1}$, consistent with the values reported by \citet{shimizu_etal_09}. 
This is more than what \cite{leka1997} finds on average, 
most probably, since our spatial resolution is significantly better. 

As described in Paper 1, a chain of elongated segments forms a bright lane within a 
LB. That these segments are smaller than granules in the quiet Sun is expected to be 
the result of the magnetic field that alters the 
mode of convection \citep[e.g.,][]{rimmele97, schussler_vogler_06, bharti_etal_09, scharmer_09}. 
The mere existence of these bright lanes provides evidence of overturning magnetoconvection in the LB.

During the sequence of our observation, the LB area becomes smaller and observations from July 6 show that the 
LBs have disappeared. This means that the initially weakly magnetized plasma gradually becomes magnetized and the field strength 
increases, thereby further suppressing the regular type of convection. In other words, the magnetic field 
gradually intrudes into the gap of weak field. During this evolution the LB transforms into 
an area of umbral dots and diffused emission, and finally fades away into an umbral area. 

\subsection{Light bridges as ignitor for penumbra formation} \label{sec:ignitor}

Although LBs fade away as the spot develops, they seem to play a crucial role in the penumbra formation process.  
We find that the first penumbral filaments form in the vicinity of LBs. This agrees with observations 
by \cite{yang_etal03}.  There are more examples where the first penumbral filaments form as extensions of LBs,  
as seen in the DOT archive \citep[][e.g., spot of July 13, 2005]{dot}. 

However, this property does not apply to areas facing the opposite AR polarity. 
No stable penumbra form on these ends of the LBs. 
We ascribe this to the ongoing activity and surmise that the presence of magnetic field in 
the emergence site hinders the  magnetic field of the protospot to spread out.

On the side opposite to the emergence site, the magnetic field has space to unfold, 
to become more inclined and to form a penumbra. 
That this happens at the ends of LBs could indicate that the transition in 
the penumbral mode of convection starts where the field strength is reduced. 
The effect of granular motions onto the magnetic fields should be strongest at locations of the weak magnetic field, 
and it seems plausible that this should be the locations where the penumbra forms first. 
However, from our observation we cannot characterize this transition further. 

It has been suggested by \cite{weiss_etal_2004} that field lines are dragged downwards by granular motions \citep[see also][]{wentzel92}. 
Thereby the field inclination increases and a penumbra forms. In their scenario this happens at corrugations 
of the magnetopause, which are caused by the fluting instability \citep[e.g.,][]{schuessler_84}. Inspired by our 
observation, we speculate that such a process would preferably occur where the field is weak, i.e. only 
at the ends of LBs. There it could also be the fluting instability that initiates the transition from LB 
into penumbra, but neither our observations nor the model calculations of \cite{weiss_etal_2004} allow further insight.

\subsection{Light bridges as channels for magnetic flux} \label{sec:channel}

The LBs may yet have an additional key-role in transforming a proto-spot without penumbra into a spot with penumbra: 
They are the channels in which the magnetic flux is transported from the one side of the spot to the other. 
This results from the following considerations:

From the temporal evolution we see that SFPs that migrate towards the spot tend to merge the spot near the right end 
of the upper LB (cf. Fig.\ref{fig:table_maps}).  We have presented evidence that the increase in magnetic flux of the 
evolving spot comes from the flux that is added by SFPs. In this sense, LBs seem to be the entry gate for a new 
flux that merges the spot. Since the penumbra forms on the other side, while the magnetic neutral line relative 
to the umbra does not change its location, magnetic flux must be transported through the spot, from one side to 
the other. And since the new flux enters the spot in the LB and the penumbra starts to form on the other side of 
the LB, it seems plausible to assume that the flux is transported through the LB and not through the areas of strong fields.
We envisage two processes that could accomplish such a transport of flux: 

\paragraph{(1) Advection of flux:} 
In the first possibility, a stream of plasma advects the flux elements from one side to the other. 
This seems possible because it would take some two hours to cross the LB if the flow has a speed of 
1\,km\,s$^{-1}$ \citep{hirzberger_etal_2002,louis_etal_08}. Visual inspection of our G-band movie supports 
the presence of such a flow:  Proper motions are seen that could be a signature of shear flows in the LB \citep{kneer_etal_2011}. 

\paragraph{(2) Reconfiguration:} For this process we assume that the spot is in quasi-stationary equilibrium 
at all times during its growth phase, since the dynamical timescale is less than one hour.  
We envisage that the spot evolves through a sequence of quasi-stationary equilibria.  
When new magnetic flux joins the spot (a SFP merges), the magnetic field configuration is perturbed 
and the spot is out of balance. This perturbation causes a perturbation in magnetic field strength 
and hence in magnetic pressure. This pressure perturbation, which consists of a compression of field lines, 
propagates through the spot with Alfv\'en speed and brings the spot back into magnetostatic equilibrium. 
In this way, magnetic flux is redistributed within the spot. 
Thereby, this process could explain the increase in magnetic 
flux beyond the magnetic neutral line. These waves of magnetic pressure variations would preferably propagate through the LBs, 
since there the field strength is lower and less energy is needed for the field compression than in the umbra.

If one such a compression wave reaches the other side, the spot has grown in magnetic flux and is in 
a new equilibrium. In this new configuration the outermost field lines are more inclined, which potentially 
favors the formation of a penumbra.

Both these processes could exist in parallel, and both would favor the end of LBs as the locations where the 
formation of the penumbra starts. In the further evolution, when the flow through the LBs has ceased and when 
the field strength in LB has increases, only the `reconfiguration' process remains. Then the pressure perturbations 
travel through the umbra, and the segments where no penumbra has yet formed can be filled.

\subsection{Some properties of the developing penumbra}\label{sec:developed}

Wherever and whenever continuum and flow maps show the signatures of a penumbra, the magnetic field has 
the typical properties of a penumbra: A strength of some 1.5\,kG and a mean inclination of some 50$^\circ$. 
However, at the earliest stages, the magnetic area of the protospot extends over the visible limits. 
There the field strength is also more than 1\,kG, and the field has large inclinations, 
but neither continuum maps nor the flow maps show a sign of a penumbra. During the course of our 
time series, these areas are mostly converted into penumbral areas. Unfortunately, we have no explanation for why there 
is flux in the early stages on the side opposite to the emergence site, which is not yet integrated into 
the visible boundaries of the spot.

In Sect.~\ref{sec:curved} we report on the presence of opposite polarities at the outer end of the curved filaments.
The presence of these opposite polarities during the formation could be interpreted as the 
turbulent pumping process that is described in \citet{weiss_etal_2004}: field lines that reach 
some critical angle of inclination, which are grabbed by convective plumes and dragged downwards. 
This could explain the opposite polarity at the boundary of the spot as the footpoints of returning field lines.

\subsection{When does a penumbra form?}

Last but not least, we want to discuss the question about the conditions that are needed to form a penumbra. 
The first condition is a minimum of magnetic flux. According to \cite{zwaan87}, a flux of at least about $5\times\!10^{20}$\,Mx 
is required to assemble  a spot. This is consistent with the result of \cite{bogdan_etal_88} who note that the minimum umbral 
radius in a large sample of spots is about 1.5\,Mm (assuming that the radial extension of the penumbra is that 
of the umbra and plugging in typical values for the field strength).
\cite{leka_sku98} observed a sunspot that had a partial penumbra with a magnetic flux of $1\times\!10^{20}$\,Mx, 
comparable to critical flux of $1-7\times\!10^{20}$\,Mx reported by \cite{rucklidge_etal1995}. 
Therefore, various authors come to the comparable results for the critical flux:  
$\phi_{\mathrm crit}$ \,$\lesssim\!5\times\!10^{20}$\,Mx. The protospot in our data is clearly larger with a 
magnetic flux of 17\,$\times\,10^{20}$\,Mx. 

In addition, the properties of the magnetic field certainly pose necessary conditions: As seen in the top rows 
of Fig.~\ref{fig:table_maps}, we find for the critical magnetic field strength, $B_{\mathrm crit} \lesssim1.6$\,kG. 
From a comparison of the maps of continuum (Fig.~\ref{fig:table_maps}) and inclination (Fig.~\ref{fig:local}), 
we find that the existence of a penumbral filament is associated with large inclination 
angles, $\alpha_{\mathrm crit} \gtrsim 60^{\circ}$. These critical values imply that the spot is growing, 
and its flux is increasing such that the field strength and inclination become critical.

Moreover, we find that stable filaments only form away from the emergence site. 
Therefore, it seems necessary to have a `quiet' environment into which the penumbra can grow. 
These are conditions that are fulfilled in the forming penumbra of our data set, 
but we cannot tell whether they are sufficient.

\section{Conclusion}
\label{sec:conclusion}

We present the spectropolarimetric data of an emerging active region and a forming penumbra. The emergence site is characterized by 
elongated granules that are associated with small flux patches (SFPs) of opposite polarities. These SFPs coalesce to form pores 
and spots. Spots partly form from merging pores, but a substantial fraction of the flux also comes from merging SFPs. 
As a result, SFPs are the building blocks of structure formation in ARs.

During the forming phase, LBs are the remnants of merging pores and they play a key role in the formation process.  
LBs provide the channels through which the flux can be transported from the emergence side to the other side of the spot, where 
the penumbra forms. This transport could be accomplished ($a$) by a flow that advects the flux or ($b$) by readjustments 
of the magnetic field configuration to accommodate the increasing flux, and the associated waves of magnetic pressure 
variation favor propagating in LBs. At the same time, LBs are the ignitors for the formation of penumbra filaments as 
the penumbra starts to form next to the ends of LBs.

The formation of the penumbra is linked to a set of conditions as there are critical values for the magnetic flux of the protospot, 
for the field strength, and for the inclination angle of the field. In presence of these conditions and 
a `quiet' magnetic surrounding, individual stable filaments form on a dynamical timescale.

\begin{acknowledgements}
The German VTT is operated by the
Kiepenheuer-Institut f\"ur Sonnenphysik at the Spanish Observatorio del Teide. 
The success of the observational setup was achieved by the common effort of members of 
the institutes involved in the development of the German solar observatories (IAP, IAC, MPS, KIS). 
We are grateful to Bruce Lites for assisting us with the AZAM code. 
We wish to thank Christian Beck for helpful suggestions. 
NBG acknowledges financial support by the DFG grant Schm 1168/9-1. 
R.S. acknowledges fruitful discussions at the workshop on ``Filamentary Structure and Dynamics of 
Solar Magnetic Fields'' at the ISSI in Bern.
\end{acknowledgements} 

\bibliography{rezabib}
\end{document}